\documentclass{ws-ijmpa}
\usepackage{epsfig}
\begin{document}

\markboth{Pyungwon Ko}{Electroweak symmetry breaking  and
cold dark matter from hidden sector technicolor}

\catchline{}{}{}{}{}
%
%%%%%%%%%%%%%%%%%%%%%%%%%%%%%%%%%%%%%%%%%%%%%%%%%%%%%%%%%%%%%%%%%%%%

\title{Electroweak symmetry breaking
and cold dark matter  from hidden sector technicolors
%\footnote{For the title, try not to use more than 3 lines. Typeset the title in
%10 pt roman, uppercase and boldface.}
 }

\author{PYUNGWON KO
%\footnote{
%Typeset names in 8 pt roman, uppercase. Use the footnote to indicate the
%present or permanent address of the author.}
}

\address{School of Physics, Korea Institute for Advanced Study \\
Seoul 130-722, Korea
%\footnote{State completely without
%abbreviations, the affiliation and mailing address, including
%country. Typeset in 8 pt
%italic.}
\\
pko@kias.re.kr}

%\author{Jae Yong Lee
%\footnote{
%Typeset names in 8 pt roman, uppercase. Use the footnote to indicate the
%present or permanent address of the author.}
%}

%\address{School of Physics, Korea Institute for Advanced Study \\
%Seoul 130-722, Korea
%\footnote{State completely without
%abbreviations, the affiliation and mailing address, including
%country. Typeset in 8 pt
%italic.}
%\\
%littlehigg@kias.re.kr}

%\author{SECOND AUTHOR}
%
%\address{Group, Laboratory, Address\\
%City, State ZIP/Zone, Country\\
%second\_author@domain\_name}

\maketitle

\begin{history}
\received{Day Month Year}
\revised{Day Month Year}
\end{history}

\begin{abstract}
We consider models with  a vectorlike confining gauge theory in
the hidden sector, and demonstrate that the origin of the
electroweak symmetry breaking (EWSB) is due to the dimensional
transmutation in the hidden sector gauge theory, and the lightest
mesons in the hidden sector could be a good cold dark matter (CDM)
candidate. There would be more than one neutral Higgs-like scalar
bosons, and they could decay mainly into the CDM pair, if  that
decay channel is kinemtically allowed. 
\end{abstract}

\ccode{PACS numbers: }

\section{Introduction}

Revealing the origin of the electroweak symmtry breaking (EWSB) is
the most pressing question in particle physics in the era of CERN
Large Hadron Collider (LHC). Another important problem in particle
astrophysics and cosmology is to identify the nature of cold dark
matter (CDM). Also there is a more speculative issue about the
existence of a new hidden sector, which is generic in
supersymmetric (SUSY) model buildings or superstring theories.

In this talk, I would like to consider three seemingly unrelated
questions:
\begin{itemize}
\item Can all the masses arise (mostly) from quantum mechanics, as
in massless QCD ?

\item What is the nature of CDM ? Is it possible to have all the
global symmetry as accidental symmetries, as in the standard model
(SM) ?

\item What would be the phenomenological consequences, if there is
a hidden sector ?
\end{itemize}
I will present models with a hidden sector where these seemingly
unrelated questions are in fact closely connected with each other.
More details and complete list of references can be found in
Ref.s~ \refcite{ko1,ko2}.
%\cite{ko2}.

Let me remind you that there is  a good old example, namely
quantum chromodynamics(QCD), where we can learn many lessons
related with the issues listed above. QCD has many nice features:
renormalizability, asymptotic freedom, confinement and chiral
symmetry breaking, dynamical generation of hadron masses, natural
hierarchy between the Planck scale and the QCD scale $\Lambda_{\rm
QCD}$. In addition pions are stable if electroweak interactions
are switched off. It would be nice if we could have a model for
EWSB in the same manner as the dimensional transmutation in QCD,
and CDM is stable as pions are stable under strong interaction.

The basic features of our models are the following. We assume a
vectorlike confining gauge theory such as QCD or technicolor in
the hidden sector, which we dub as hidden sector technicolor
(hTC). Then dimensional transmutation will occur in the hidden
sector, and this scale is transmitted to the SM by a messenger,
and triggers EWSB. And the lightest mesons in the hidden sector
becomes a CDM.

\section{Model I}

Let us assume that there is a new strong interaction that is described
by $SU(N_{h,C})$ guage theory with vectorlike quarks ${\cal Q}_i$ and
$\overline{\cal Q}_i$ with $N_{h,f}$ flavors, such as QCD with the
confinement scale $\Lambda_h$.
This scale is presumed to be higher than the electroweak scale by at least
an order of magnitude.
\begin{equation}
{\cal L}_{hidden} = - {1\over 4}{\cal G}_{\mu\nu} {\cal G}^{\mu\nu} +
\sum_{k=1}^{N_{HF}}
\overline{\cal Q}_k ( i {\cal D} \cdot \gamma - M_k ) {\cal Q}_k
\end{equation}
Then this new strong interaction will trigger chiral symmetry
breaking due to nonzero $\langle {\cal Q \overline{Q}} \rangle
\equiv \Lambda_{H,\chi}^3$. For illustration, we assume that there
is an approximate $SU(2)_L \times SU(2)_R$ global symmetry in the
hidden sector that breaks down to $SU(2)_V$ spontaneously.
%Below the chiral symmetry breaking scale,
%there appear mesons and baryons made of ${\cal Q}$'s and
%$\overline{\cal Q}$'s. The lightest mesons are massless
%Nambu--Goldstone bosons, and the lightest baryons are analogous to
%ordinary nucleons, with mass $M_{h,N} \simeq N_h \Lambda_h$.
In the low energy limit of hTC,  massless Nambu-Goldstone bosons
will appear, which are dubbed as hidden sector pion $\pi_h$. Also
there would be a scalar resonance like the ordinary $\sigma$, and
we call it $\sigma_h$, and $\vec{\pi}_h$ and $\sigma_h$ will form
$SU(2)_L \times SU(2)_R$ bidoublet (denoted as $H_2$) and the low
energy effective theory will be the same as the Gelmann-Levy's
linear $\sigma$ model, except that the mesons are in the hidden
sector, so that SM singlets. They are all neutral.

The potential for the SM Higgs and the hidden sector $H_2$ is given by
\begin{eqnarray}
V ( H_1 , H_2 ) & = & -\mu^2_1 (H^\dagger_1
H_1)+\frac{\lambda_1}{2} (H^\dagger_1 H_1)^2 -\mu^2_2 (H^\dagger_2
H_2) \nonumber + \frac{\lambda_2}{2} (H^\dagger_2 H_2)^2
\nonumber
\\
& + & \lambda_3 (H^\dagger_1 H_1)(H^\dagger_2 H_2) + \frac{a
v^3_2}{2}\sigma_h
\end{eqnarray}
This looks like the potential in the 2-Higgs doublet model, but
there are important differences. First, $H_2$ is a SM singlet, not
a SM doublet. $W$ and $Z^0$ get masses entirely from $H_1$ VEV.
And the $a$ term is new in our model, and necessary to generate
the mass for the hidden sector pion. Note that the $\lambda_3$
term connects the SM and the hidden sector, and originates from
nonrenormalizable interactions between two sectors, or by some
messengers.

It is straightforward to analyze phenomenology from this scalar
potential. The generic predictions of our models are the
following:

\begin{itemize}
\item The origin of the EWSB, namely the negative Higgs mass$^2$
parameter could be the chiral symmetry breaking in the hTC.

\item The electroweak precision test does not put strong
constraints unlike in the ordinary technicolor models, since $H_2$
does not contribute to the $W$ and $Z^0$ masses at tree level. And
no Higgs-mediated FCNC problem since $H_2$ does not couple to the
SM fermions.

\item There are more than one neutral Higgs-like scalar bosons,
and they can decay into the $\pi_h$ with a large invisible
branching ratio. This makes relatively difficult to produce and
discover these Higgs-lilke neutral scalars at colliders. See Fig.
1 (a) and (b).

\item The hidden sector pion ($\pi_h$) is stable due to the flavor
conservation in the hTC, and could be a good CDM candidate. Direct
detection rate of the $\pi_h$ is in a promising sensitivity of the
current/future DM detection experiments such as CDMS, XENON10 or
XMASS (Fig.~2 (a)).

%\item The Model-II has a classical
%scale invariance, and all the mass scales are generated by quantum
%mechanics.
\end{itemize}

\begin{figure}
\includegraphics[width=6cm] {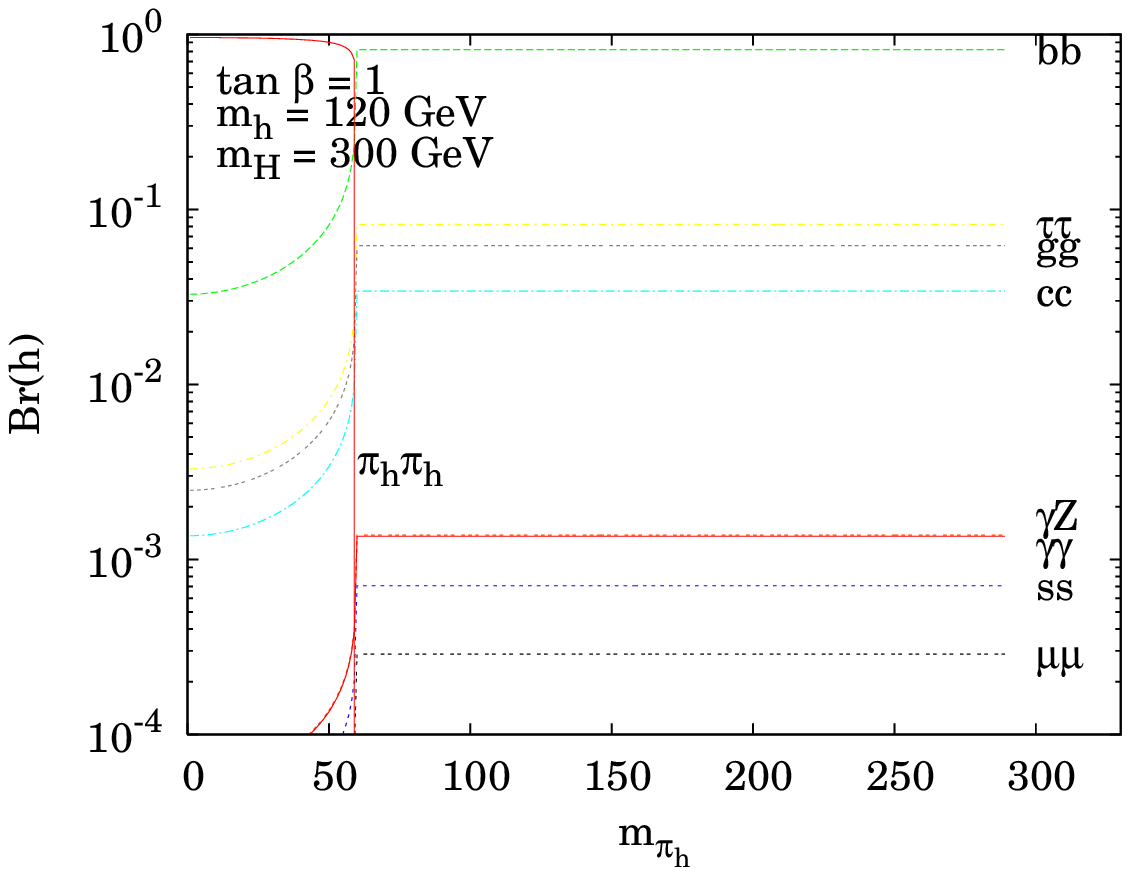}
\includegraphics[width=6cm]{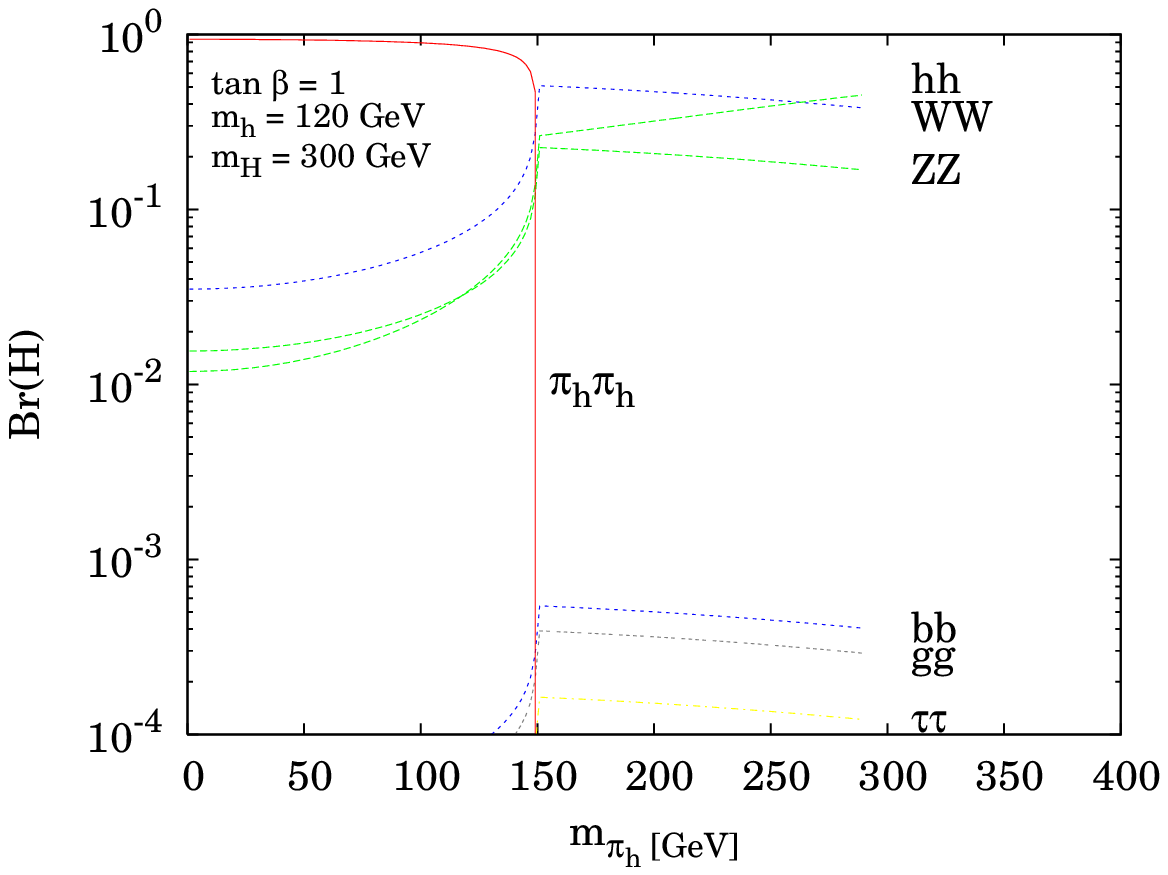} % {hh120-300.eps}
%\includegraphics[width=7cm] {h120-500-1-2.eps}
%\includegraphics[width=7cm] {hh120-500-1-2.eps}
%\vspace{-5mm}
\caption{\label{fig:br-t1} Branching ratios of (a) $h$ and (b) $H$
as functions of $m_{\pi_h}$ for $\tan\beta = 1$, $m_h = 120$ GeV
and $m_H = 300$ GeV.}
\end{figure}

%\begin{figure}
%\includegraphics[width=6cm] {cont-500-1-k1.eps}
%\includegraphics[width=6cm] {cont-500-1-k2.eps}
%%\vspace{-5mm}
%\caption{ $\Omega_{\pi_h} h^2 $ in the $( m_{ h_1 } , m_{\pi_h} )$
%plane for $\tan\beta = 1$ and $m_H = 500$ GeV.}
%\end{figure}

\begin{figure}
\includegraphics[width=6cm] {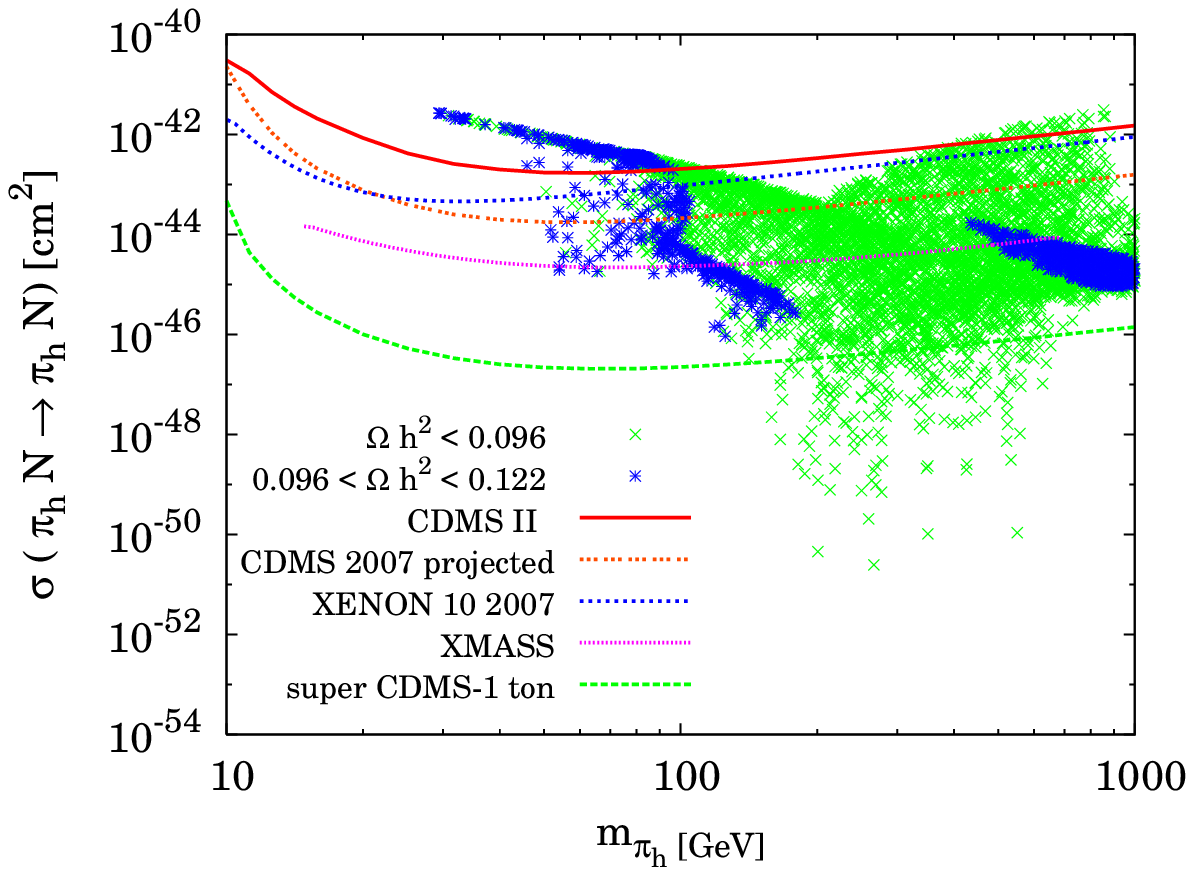}
\includegraphics[width=6cm]
{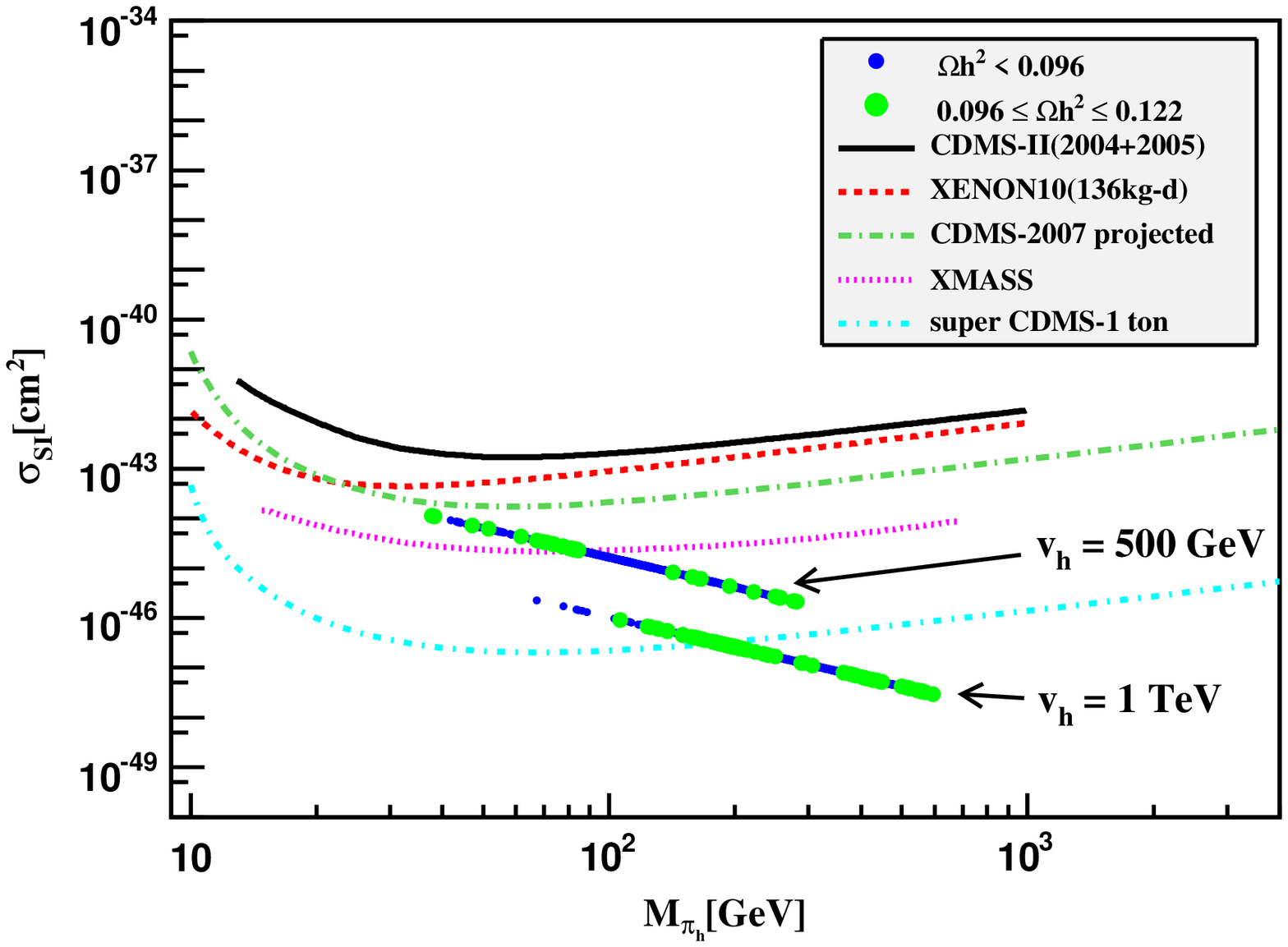}
%\includegraphics[width=6cm] {dr-5.eps}
%\vspace{-5mm}
\caption{\label{fig2} $\sigma_{SI} (\pi_h p \rightarrow \pi_h p )$
as functions of $m_{\pi_h}$ for (a) $\tan\beta = 1$ in Model I,
and (b) Model II.
%, with thee upper one is for $v_h = 500$ GeV and
%$\tan\beta = 1$, and the lower one is for $v_h = 1$ TeV and
%$\tan\beta = 2$..}
}
\end{figure}

\section{Model II with classical scale invariance}

The Model I has a few drawbacks, since the hidden sector quark
masses $M_k$'s are given by hand, and the Model I is not
renormalizable. These can be cured by introducing a real singlet
scalar $S$ and making the following replacement, $M_k \rightarrow
\lambda_k S$ in Eq. (1). Then ${\cal L}_{\rm hidden}$ has
classical scale symmetry. With a real singlet $S$, the SM
lagrangian is implemented into
\begin{equation}
  \label{eq:sm}
  {\cal L}_{\rm SM}  =  {\cal L}_{\rm kin} + {\cal L}_{\rm Yukawa}
- {\lambda_{H} \over 4}~( H^{\dagger} H )^2 - {\lambda_{SH} \over
2}~S^2 ~ H^{\dagger} H - {\lambda_S \over 4}~S^4
\end{equation}
assuming classical scale symmetry. Since there are no mass
parameters in this lagrangain, this is a suitable starting point
to investigate if it is possible to have all the masses from
quantum mechanical effects. Note that the real singlet scalar $S$
plays the role of messenger connecting the SM Higgs sector and the
hidden sector quarks.

Dimensional transmutation in the hidden sector will generate the
hidden QCD scale and chiral symmetry breaking with developing
nonzero $\langle \bar{\cal Q}_k {\cal Q}_k \rangle$. Then the
$\lambda_k S $ term generate the linear potential for the real
singlet $S$, leading to nonzero $\langle S \rangle$. This in turn
generates the hidden sector current quark masses through
$\lambda_k$ terms as well as the EWSB through $\lambda_{SH}$ term.
The $\pi_h$ will get nonzero masses, and becomes a good CDM
candidate. Due to the presence of the messenger $S$, the CDM pair
annihilation into the SM particles occurs more efficiently in
Model II than in Model I, and it is easy to accommodate the WMAP
data on $\Omega_{\rm CDM} h^2$. Direct detection rates are in the
interesting ranges (see Fig.~2 (b)). All the qualitative features
of this model is similar to the Model I. See Ref.~\refcite{ko2}
%and \refcite{ko3}
for more details.

\section{Conclusions}

In this talk, I presented models where the origin of EWSB and CDM
lie in the hidden sector technicolor interaction. In the Model II,
all the masses including the CDM mass arise quantum mechanically
from dimensional transmutation in the hidden sector. One can enjoy
many variations of these models by considering different gauge
groups and matter fields in the hidden sector. If we include the radiative
corrections to the scalar potential, the details could change,
but the qualitative features described in this talk would remain untouched.

\section*{Acknowledgments}
I thank Dr. Chun Liu for invitation to this nicely organized
conference. I am grateful to Taeil Hur, D.W. Jung and J.Y. Lee for
collaborations. This work is supported in part by KOSEF through
CHEP at Kyungpook National University.

\appendix

%\begin{thebibliography}{000} %for 3 digits
%\begin{thebibliography}{00}  %for 2 digits

\end{document}